\documentclass[pra,aps,showpacs,superscriptaddress,amsmath,amssymb,twocolumn]{revtex4}

\usepackage{graphicx}
\usepackage{dcolumn}
\usepackage{bbm}

\newcommand{\be}{\begin{equation}}
\newcommand{\ee}{\end{equation}}
\newcommand{\ba}{\begin{array}}
\newcommand{\ea}{\end{array}}
\newcommand{\bea}{\begin{eqnarray}}
\newcommand{\eea}{\end{eqnarray}}

\newcommand{\bra}[1]{\ensuremath{\langle #1 |}}
\newcommand{\ket}[1]{\ensuremath{| #1 \rangle}}

\begin{document}

\title{Controlling entanglement by direct quantum feedback}

\author{A. R. R. Carvalho}\author{A. J. S. Reid}
\affiliation{Department of Physics, Faculty of Science, The Australian National University, ACT 0200, Australia}
\author{J. J. Hope}
\affiliation{Australian Centre for Quantum-Atom Optics, Department of Physics, Faculty of Science, The Australian National University, ACT 0200, Australia}
\date{\today}

\begin{abstract}
We discuss the generation of entanglement between electronic states of two atoms in a cavity using direct quantum feedback schemes. We compare the effects of different control Hamiltonians and detection processes in the performance of entanglement production and show that the quantum-jump-based feedback proposed by us in Phys. Rev. A {\bf 76} 010301(R) (2007) can protect highly entangled states against decoherence. We provide analytical results that explain the robustness of jump feedback, and also analyse the perspectives of experimental implementation by scrutinising the effects of imperfections and approximations in our model.

\end{abstract}


\maketitle

\section{Introduction}

Recent experimental advances have enabled individual systems to be monitored and manipulated at the quantum level in real time~\cite{hood98,julsgard01,lu03,geremia04,ottl05,puppe07}. As new advances continue to emerge, and the development of control strategies for quantum systems becomes essential, quantum technology inevitably approaches the well-developed classical control theory, borrowing its concepts and extending them to the quantum realm. An important example is feedback control, which consists of the manipulation of the system according to the information acquired through measurement. Feedback control has been applied to quantum systems~\cite{belav,wise_milb93,wiseman94,doherty99}, including its implementation in a variety of experimental setups~\cite{geremia04,reiner04,morrow02,bushev06}.

Relying on the ability to produce specific states and perform controlled operations on them, quantum information is undoubtedly an area that would benefit from further advances in quantum control. Of particular interest in this context is the preparation of entangled states, indispensable for quantum information processing. Notwithstanding the considerable number of experiments carried out on the generation of entanglement~\cite{bouwmeester99,rauschenbeutel00,sackett00, panPRL01, roos}, and the effort to protect the system against undesirable imperfections and interactions, entanglement decay due to uncontrolled coupling with the environment remains a major problem yet to overcome~\cite{roosPRL04,eberly_04,arrc_mpd}. In this scenario, quantum feedback emerges as a possible route to develop strategies to circumvent entanglement deterioration. 

In fact, quantum feedback control has been recently used to improve the creation of steady state entanglement in open quantum systems. A Markovian, or direct, feedback~\cite{wise_milb93,wiseman94} was used in~\cite{mancini05,wang05} to show that states with entanglement corresponding to one third of the maximum possible value can be produced. Maximally entangled states could be achieved~\cite{stockton04} in an idealised situation, by the use of Bayesian~\cite{doherty99}, or state estimation, feedback. This improvement, however, comes at the cost of an increasing experimental complexity due to the need of a real time estimation of the quantum state in the later method, as compared to the simple feedback directly proportional to the measurement signal proposed by the former strategy. 

Despite being simpler to implement, direct feedback still exhibits a multitude of possibilities due to the arbitrariness of choices for the control Hamiltonian and measurement schemes. In a recent Rapid Communication~\cite{jumpfeedback}, we have shown that an appropriate selection of the feedback Hamilton and detection strategy leads to the robust production of highly entangled states of two atoms in a cavity. In this contribution, we will further explore the richness of feedback strategies by comparing the jump-based scheme proposed in~\cite{jumpfeedback} with a strategy based on homodyne measurements~\cite{wang05}, taking into account the unavoidable imperfections that may occur in realistic experimental implementations.

The paper is organized as follows. In the first part of Section~\ref{sec:model} we introduce the model of the system and its equation in the absence of feedback. In the second part we describe the introduction of feedback and the corresponding changes in the master equation for two different choices of measurement strategy, namely, photodetection and homodyning. Entanglement dynamics is examined in Section~\ref{sec:entang}, and the role of feedback on entanglement generation is analysed. Robustness issues are discussed in Section~\ref{sec:robust}, where the effects of detection inefficiencies and spontaneous emission on steady-state entanglement are considered. Section~\ref{sec:exp} discusses different experimental regimes and explores the non-adiabatic limit of the model. Section~\ref{sec:conc} concludes with a discussion of our main results.

\section{The model}\label{sec:model}

\begin{figure}
\includegraphics[width=6.0cm]{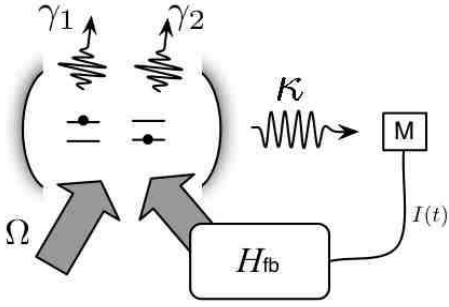}
\caption{Schematic view of the model. The system consists of a pair of two-level atoms coupled to cavity, which is driven by a laser field and damped. Conditioned on the measurement of the output of the leaky cavity, a Hamiltonian is applied to the atoms, completing the feedback scheme.} 
\label{fig1}
\end{figure}

The system consists of a pair of two-level atoms coupled resonantly to a single cavity mode, with equal coupling strength $g$, and simultaneously driven by a laser field with Rabi frequency $\Omega$. 
The cavity mode is damped with decay rate $\kappa$ and the atoms can spontaneously decay with rates $\gamma_1$ and $\gamma_2$. Feedback is applied conditioned on the measurement of photons leaving the cavity, as schematically shown in Fig.~\ref{fig1}.
In the absence of feedback, the master equation describing the system is given by
\bea
\dot \rho=-i \Omega \left[(J_+ + J_-),\rho \right] -i g \left[(J_+ a + J_- a^\dagger),\rho \right] \nonumber \\  + \kappa {\cal D}[a]\rho + \sum_i\gamma_i{\cal D}[\sigma_i]\rho.
\label{eq:nofb_full}
\eea
The first and second terms represent the Hamiltonian evolution induced by the laser driving and atom-cavity coupling, respectively. The superoperator
\be
{\cal D}[c]\rho \equiv c \rho c^{\dagger}-\frac{1}{2}\left(c^{\dagger}c \rho+\rho c^{\dagger}c\right)
\ee
describes cavity and atomic decays, given in terms of $a$, the annihilation operator of photons in the cavity, and $\sigma_i=\ket{g_i}\bra{e_i}$, the lowering operator for the $i$-th two level atom, respectively. The angular momentum operators are defined as 
\bea
\label{J_operators}
J_{-}=\sigma_1+\sigma_2,\\
J_+=\sigma_1^{+}+\sigma_2^{+},
\eea
with $\sigma_i^{+}=\ket{e_i}\bra{g_i}$ the raising operator for the atomic electronic levels.

In the limit where the cavity decay rate $\kappa$ is much larger than the other relevant frequencies of the problem, the cavity mode can be adiabatically eliminated and one obtains a master equation just for the atomic degrees of freedom~\cite{wang05}
\be\label{eq:me_total}
\dot \rho=-i \Omega \left[(J_+ + J_-),\rho \right] + \Gamma {\cal D}[J_-]\rho + \sum_i \gamma_i{\cal D}[\sigma_i]\rho,
\ee
where $\Gamma=g^2/\kappa$ is the effective collective decay rate. Furthermore, the Dicke model~\cite{agarwal74} can be recovered under the assumption that the collective decay rate is much larger than the
spontaneous emission rates, $\Gamma \gg \gamma_1,\, \gamma_2$,
\be
\label{eq:me_dicke}
\dot \rho= {\cal L} \rho=-i\Omega\left[(J_+ + J_-),\rho \right] + \Gamma {\cal D}[J_-]\rho.
\ee
Later we shall relax the approximations used to obtain Eqs.~(\ref{eq:me_total}) and~(\ref{eq:me_dicke}): The non-adiabatic regime is important when considering experimental situations where high quality cavities are used, so it will be discussed in Section~\ref{sec:exp}. Section~\ref{sec:robust} will show that even small spontaneous emission rates can have a drastic deleterious effect on the final amount of entanglement produced in this system. For the moment, Eq.~(\ref{eq:me_dicke}) will be the starting point to introduce the description of the measurement scheme and the feedback mechanism.

Here we will consider the direct, or Markovian, feedback introduced by Wiseman and Milburn~\cite{wise_milb93,wiseman94}, where the control Hamiltonian is proportional to the measurement signal. This control mechanism has the advantage of being simple to apply in practice, since it avoids the challenge of real time state estimation required in Bayesian, or state-based, feedback~\cite{doherty99}. The idea is depicted in Fig.~\ref{fig1}: the cavity output is measured by a detection scheme $M$ whose signal $I(t)$ provides the input to the application of the control Hamiltonian $H_{\rm fb}=I(t)\,F$. Here, we will consider the measurement stage $M$ to be either a homodyne or a direct photodetection of the output field. 

In the homodyne-based scheme, the detector registers a continuous photocurrent, and the feedback Hamiltonian is constantly applied to the system. Conversely, in the photocounting-based strategy, the absence of signal predominates and the control is only triggered after a detection click, {\it i.e.} a quantum jump, occurs. This is reflected in the different forms of the equations representing the dynamics of a feedback-controlled system under either of the measurement schemes. In the homodyne case, Eq.~(\ref{eq:me_dicke}) becomes~\cite{wise_milb93,wiseman94}
\bea
\label{eq:hd}
\dot \rho= {\cal L}_h \rho=-\frac{i}{\hbar}\Omega\left[(J_+ + J_-),\rho \right] + \Gamma {\cal D}[J_-]\rho+ \frac{1}{\Gamma}{\cal D}[F]\rho \nonumber \\-\frac{i}{\hbar}\left[F,-iJ_-\rho+i\rho J_+\right].
\eea
This equation was used by  Wang, Wiseman and Milburn (WWM)~\cite{wang05} to show that the amount of steady state entanglement in this atom-cavity model can be enhanced above that generated by the uncontrolled dynamics~\cite{schneider02}. More recently, the same equation was used to explore the effect of different feedback Hamiltonians on entanglement generation~\cite{li08}.

In a measurement scenario based on photodetections, the unconditioned dynamics, including feedback, reads~\cite{wiseman94} 
\be
\label{eq:pd}
\dot \rho={\cal L}_p \rho=-\frac{i}{\hbar}\Omega\left[(J_+ + J_-),\rho \right] + \Gamma {\cal D}[U J_-]\rho.
\ee
The manifestation of the abrupt character of the jump feedback is clear when one writes the last term of Eq.~(\ref{eq:pd}) explicitly: ${\cal D}[U J_-]= U J_- \rho J_+ U^{\dagger}-(J_+ J_- \rho+\rho J_+ J_-)/2$. The unitary transformation $U=\exp\left[-i F \delta t/\hbar \right]$, representing the finite amount of evolution imposed by the control Hamiltonian on the system, only acts immediately after a detection event, which is described by the first term of the superoperator ${\cal D}$.
Note that this is equivalent to replacing the jump operator $J_-$ in the master equation by $UJ_-$. In this way, one can engineer the reservoir dynamics~\cite{poyatos,arrc_sp} via the transformation $U$ to generate highly entangled states.

From the differences between Eqs.~(\ref{eq:hd}) and (\ref{eq:pd}) it is evident that the resulting dynamics should strongly depend on the choice of measurement strategy. Here, we will explore this fact, together with the possibility of changing the feedback Hamiltonian to show how they affect the dynamics of the system and, in particular, its asymptotic entanglement.

\section{Steady state entanglement and feedback}
\label{sec:entang}

\subsection{Dynamics without feedback}
\label{sec:nocontrol}

The steady state solutions for the Dicke model, Eq.~(\ref{eq:me_dicke}), were obtained already many years ago~\cite{puri79,drummond80}. However, only recently their entanglement properties were brought to attention~\cite{schneider02} and explored in the context of quantum information.

The first important feature of Eq.~(\ref{eq:me_dicke}) is the fact that it is symmetric with respect to exchange of the atoms. This suggests that, instead of using the two qubit basis $\{ \ket{gg},\,\ket{ge},\,\ket{eg},\,\ket{ee}\}$, one should use angular momentum
states, and analyse the system in terms of the symmetric ($j=1$)
\be
\label{symm}
\ket{1}=\ket{gg}, \:\:\: \ket{2}=\frac{\ket{ge}+\ket{eg}}{\sqrt{2}}, \:\:\: \ket{3}=\ket{ee},
\ee
and anti-symmetric ($j=0$)
\be
\label{asymm}
\ket{4}=\frac{\ket{ge}-\ket{eg}}{\sqrt{2}}
\ee
subspaces.

The anti-symmetric subspace is a decoherence-free subspace~\cite{lidar98,zanardi97} and the state $\ket{4}$ is therefore a steady state solution of Eq.~(\ref{eq:me_dicke}). This is an interesting case, despite its trivial dynamics, since the asymptotic state in this subspace is a pure, maximally entangled, one. In fact, this situation was explored in a recent proposal for producing Werner states in a system of atoms inside a cavity~\cite{agarwal06} and in a probabilistic scheme to generate the singlet state via quantum-jump detection~\cite{plenio99}. 
In terms of dynamics, however, the symmetric case is more interesting: while in the anti-symmetric subspace an initially prepared Bell state $\ket{4}$ does not evolve at all, entanglement can be dynamically generated from any symmetrical initial condition, even from initially separable states. However, even for optimal parameters, the amount of entanglement in this case is only about $10 \%$ of the Bell state's value~\cite{schneider02}.

\begin{figure}
\includegraphics[width=9.0cm]{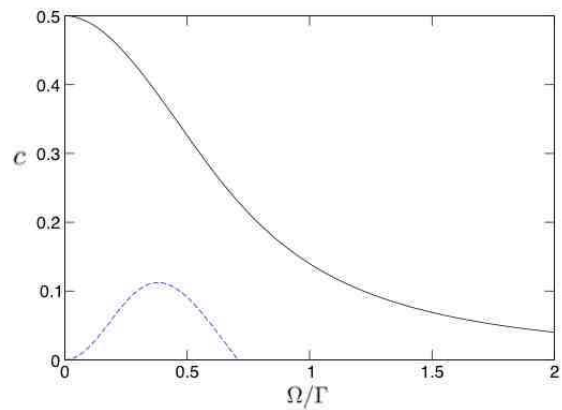}
\caption{Steady state concurrence as a function of the ratio $\Omega/\Gamma$ for initial the states $\ket{ge}$ (solid line) and $\ket{gg}$ (dashed line). In the anti-symmetric subspace there is no evolution and the system is always in the maximally entangled state $\ket{4}$ (not shown), while for the symmetric part the maximum entanglement is reached for $\Omega/\Gamma \approx 0.38$~\cite{schneider02}. The steady state for the asymmetric initial condition strongly depends on the interplay between symmetric and anti-symmetric solutions.} 
\label{fig2}
\end{figure}

For a general asymmetric initial condition the situation gets more complicated as both symmetric and anti-symmetric components are present. This is exemplified in Fig.~\ref{fig2} where the steady state entanglement, measured by the concurrence~\cite{wot98}, as a function of the ratio $\Omega/\Gamma$ is shown for an asymmetric ($\ket{ge}$, solid line) and a symmetric ($\ket{gg}$, dashed line) initial states. For zero driving, the symmetric part of the asymmetric state evolves towards the ground state ($\ket{gg}$) and the entanglement is totally due to the anti-symmetric component ($\ket{4}$). Tuning the parameters to obtain the maximum entanglement for the symmetric component ($\Omega/\Gamma \approx 0.38$) leads to a smaller amount of entanglement in the asymmetric case. This happens because in the latter case the solutions arising from different subspaces interfere, eventually resulting in a less entangled state. 

\subsection{Dynamics with feedback}
\label{sec:control}

Inclusion of Markovian feedback can significantly increase the amount of asymptotic entanglement as compared to the situation described in Fig.~\ref{fig2}. This was first shown in~\cite{wang05} for a homodyne-based feedback where the steady state entanglement was approximately 3 times larger than in the non-controlled case. Also in the case of homodyne detection, a steady-state entanglement of $80$ percent of the maximal possible value was obtained~\cite{li08} using the local asymmetric feedback law proposed in~\cite{jumpfeedback}. In this section we shall show how to improve these results by designing different feedback schemes. 

\subsubsection{The role of measurement: homodyne vs. photodetection-based feedback}
\label{sec:measure}
\begin{figure}
\includegraphics[width=8.0cm]{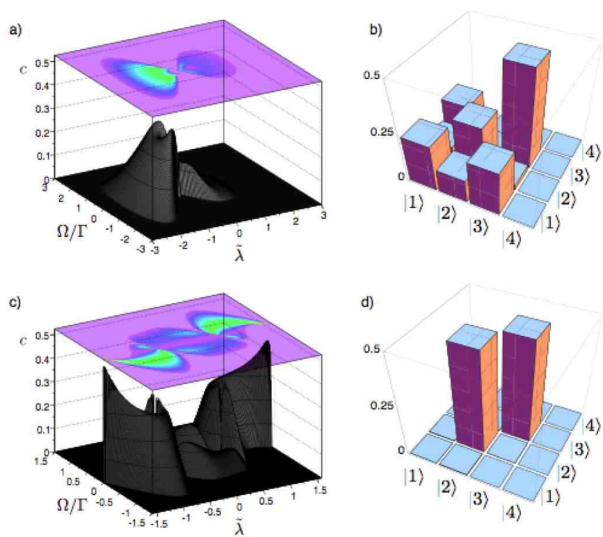}
\caption{(Color online) Steady state concurrence as a function of driving and feedback strengths (left) for homodyne (a) and photo-detection feedback (c). The homodyne case reproduces the result of~\cite{wang05} with a maximum concurrence $c\approx 0.31$. Using a jump feedback strategy, the maximum entanglement increases to $c\approx 0.49$. On the right, the absolute value of the density matrix elements (in the angular momentum basis) for the parameters corresponding to the maximum entanglement in the left panel.
} 
\label{fig3}
\end{figure}

We begin our analysis recalling the WWM results~\cite{wang05}. Their scheme fits the general scenario depicted in Fig.~\ref{fig1} when the measurement stage $M$ is a homodyne detection and the feedback Hamiltonian is $F=\lambda J_x=\lambda \left(J_-+J_+\right)$. In fact, the form of the control Hamiltonian coincides with the one appearing  in Eq.~(\ref{eq:me_dicke}) and in~\cite{wang05} was proposed to be implemented via a modulation of the field driving the cavity. With this choice of control, the final equation, Eq.~(\ref{eq:hd}), retains the symmetry properties with respect to exchange of atoms. Assuming a symmetric initial condition, the system will remain in the subspace given by Eq.~(\ref{symm}) and an analytical solution for the steady state can be found~\cite{wang05}. Figure~\ref{fig3}a shows the entanglement of these solutions as a function of the driving and feedback strengths, with a maximum concurrence $c \approx 0.3$ obtained for $\Omega/\Gamma \approx \pm 0.4$ and $\lambda/\Gamma\approx -0.8$.

Now, we change the measurement process from homodyne to photo-detection and investigate the effects of this replacement on entanglement generation. For the moment, we keep the same form for the control Hamiltonian in order to focus on the changes induced solely by the alteration of the detection scheme. Note that, as mentioned before, the feedback now is applied only when a detection occurs and is described by the action of the operator $U=\exp\left[-i F \delta t/\hbar \right]\equiv \exp[-i \tilde \lambda J_x]$. The asymptotic solution in the symmetric subspace can be calculated from Eq.~(\ref{eq:pd}) and the corresponding entanglement is shown in Fig.~\ref{fig3}c. Now, the maximum concurrence, $c\approx 0.49$, occurs at $\Omega/\Gamma \approx -0.0023$ and $\tilde \lambda/\Gamma\approx \pm -1.57$, exceeding the value obtained via homodyne-based feedback. 

The sharp differences between the two feedback schemes are evidenced by the contrasting plots of Fig.~\ref{fig3} (a) and (c) and by the form of the steady states (at the peaks) depicted in Fig.~\ref{fig3} (b) and (d), where the absolute values of the density matrix elements in the angular momentum basis, Eqs.~(\ref{symm}) and~(\ref{asymm}), are shown. In the jump feedback case, the final state at the peak concurrence is close to a mixture between $\ket{3}$ and the Bell state $\ket{2}$, the latter being the state responsible for the entanglement. In the homodyne case, the abundance of different off-diagonal elements suggests a more complicated structure. 
Understanding these differences will be important for the analysis of spontaneous emission effects in Section~\ref{sec:robust}.

\subsubsection{The role of the control Hamiltonian}
\label{sec:local}

A change in the feedback Hamiltonian to improve the control introduces an enormous range of new possibilities, even when considering the limitations imposed by constraints on experimental feasibility. In this section, however, we will restrict the discussion to the case of a local feedback, {\it i.e.} when a control Hamiltonian acting on just one of the atoms is employed, as proposed in~\cite{jumpfeedback}. The reason for this choice is that a local feedback breaks the symmetry of the system, making it possible to explore the interplay between symmetric and anti-symmetric subspaces from a dynamical point of view. Under the symmetric control described previously, all symmetric initial states lead to the same final stationary solution, while the anti-symmetric state is stationary itself. There are, however, infinitely many stationary states mixing both subspaces, which are unambiguously determined by the choice of initial conditions. The first important point to address is then how the relaxation properties of the feedback master equations change under the new asymmetric control law. 

Our discussion here will be based on the coherence-vector formalism as described by Lendi in~\cite{alicki_87}. In this approach, the original master equation is transformed into a linear system of equations in real space, {\it i.e.} $\dot \rho(t)={\cal L} \rho(t) \rightarrow {\dot{\vec v}}(t) = G \vec v(t) + \vec k$, and the stationarity properties will be encoded in the matrix $G$. A well known example of such a transformation is the description of two level systems in terms of the Bloch equations. For our purposes it suffices to note that if ${\rm det} (G) \ne 0$ then the system admits a unique stationary solution given by $\vec v_{ss}=-G^{-1} \vec k$. If ${\rm det} (G) = 0$ then two situations may arise: either the equation  $G \vec v(t) + \vec k = 0$ has no solution at all, or it has infinitely many, determined by the initial condition. 

The matrix $G$ can be straightforwardly calculated from the feedback master equations and it is obvious, from our previous discussion on the symmetric control $J_x$, that, in that case, $G$ is singular since there were more than one stationary solution. However, for a general local control of the form $U=U_1\otimes \mathbbm 1$, the matrix is non-singular (except for the trivial cases of $U_1 = \mathbbm 1$ and $\Omega = 0$) and there is a unique solution. Moreover, note that, in the case of jump-based feedback, the anti-symmetric Bell state $\ket{4}$ remains stationary for {\it any} choice of $U$ and therefore is the only steady state of the system. 

\begin{figure}
\includegraphics[width=8.0cm]{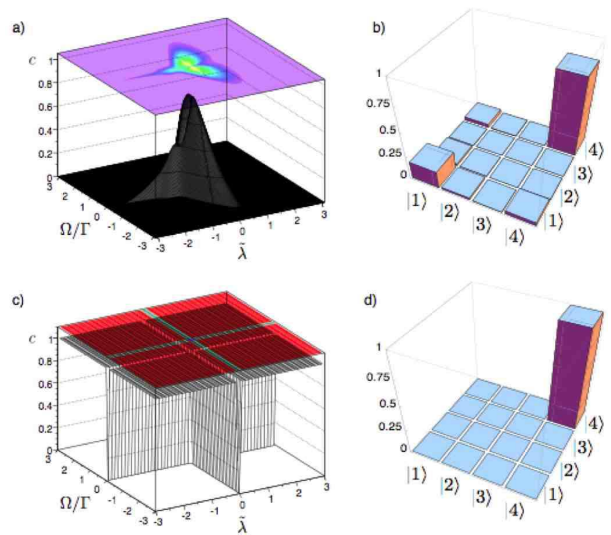}
\caption{(Color online) Steady state concurrence for a homodyne (a) and photo-detection (c) feedback with local control $\sigma_x$. There is a huge improvement as compared to the situation depicted in Fig.~\ref{fig3}. For a local control, a pure anti-symmetric Bell state is obtained for all parameters (except in the cases with zero driving $\Omega=0$ or no feedback $\tilde \lambda = 2 \pi m$), and all possible forms of $U_1$. On the right, the absolute value of the density matrix elements for the peaks on the left plots. } 
\label{figurelocal}
\end{figure}

In the case of homodyne feedback, the steady state solution is also unique, but depends on the form of $U_1$ and on the parameters $\lambda$ and $\Omega$. Figure~\ref{figurelocal} illustrates the effect of local control on the steady state entanglement for the particular choice  
\be
\label{localham}
F=\lambda \, \sigma_x  \otimes \mathbbm{1},
\ee
with $\sigma_x= \sigma_1 + \sigma_1^+$, for homodyne (a) and photo-detection (c) based feedback. In the homodyne case, there is a substantial increase in the asymptotic entanglement as compared to the $J_x$ control of Fig.~\ref{fig3}. The final state with the largest amount of entanglement corresponds now to a combination of the anti-symmetric component $\ket{4}$ with the symmetric ones, $\ket{1}$ and $\ket{2}$ (see Fig.~\ref{figurelocal}b), with $c=0.81$ for $\lambda/ \Gamma=\pm 0.01$ and $\Omega/\Gamma=\pm 0.07$. As discussed in the previous paragraph,  for a photo-detection feedback based on a local control law, a maximally entangled Bell state is generated from any initial condition for all non-trivial parameters, as shown by the plateaux in Fig.~\ref{figurelocal}c and the density matrix elements in Fig.~\ref{figurelocal}d.

Since this steady state can also be obtained directly from the non-controlled system, one could, at first sight, question the relevance of the use of feedback in this case. Indeed, if one considers the model described by Eq.~(\ref{eq:me_dicke}) without any kind of imperfection, then the jump feedback with local control would represent no advantage over the non-controlled case. Nonetheless, imperfections are inherent to any real physical system, and should be considered. Without control, for example, the asymmetric Bell state has to be produced beforehand to be unaffected by the dynamics, and any symmetric component introduced by a non-ideal preparation would spoil the final entanglement (see Section~\ref{sec:nocontrol}). Conversely, with feedback the system naturally evolves to the pure entangled state $\ket{4}$ for any initial condition. However, besides non-ideal initial state preparation, other sorts of imperfection as, for example, detection inefficiency or errors in the production of the feedback Hamiltonian, may arise from the feedback implementation, and will investigated in the next section.

\section{Robustness of control}
\label{sec:robust}
\subsection{Spontaneous emission effects}

All our previous discussions were based on the application of different kinds of feedback control on the model described by Eq.~(\ref{eq:me_dicke}), where spontaneous emission effects were neglected. However, even if all other sources of imperfection were surmounted, spontaneous emission would still be the fundamental limiting factor for the existence of entanglement in a system of atomic qubits. Consequently, the ultimate goal for the feedback schemes investigated here would be the production of steady state entanglement under a model that includes this effect. 

The starting point is therefore Eq.~(\ref{eq:me_total}). The new feedback equations are similar to Eqs.~(\ref{eq:hd}) and~(\ref{eq:pd}), differing only by the addition of the last two terms of Eq.~(\ref{eq:me_total}). The first thing to note is that those terms break the decoherence-free condition for the anti-symmetric subspace and the advantages of using feedback will become even more evident.
Analytical solutions for the steady states can still be found but, being generally cumbersome, will not be presented here in their complete form. Instead, we will analyse the regime where spontaneous emission effects can be considered as a perturbation to the system, and focus on the qualitative understanding of how these effects influence the feedback schemes discussed in the previous section. 

\begin{figure}
\includegraphics[width=8.0cm]{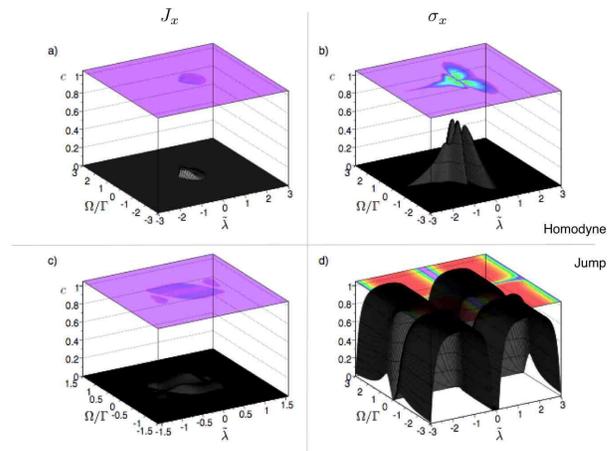}
\caption{(Color online) Spontaneous emission effects on the steady state concurrence for the cases corresponding to Fig.~\ref{fig3} (left, $J_x$ conrol) and Fig.~\ref{figurelocal} (right, $\sigma_x$ control). Even for a small decay rate ($\gamma/\Gamma=0.01$), the effects for a $J_x$ control are pronounced for both homodyne (top) and jump (bottom) detection schemes. In the local control case (right) the shape seen in Fig.~\ref{figurelocal} is left basically unaltered with only a small decrease in the maximum value of concurrence. Remarkably, for the local control Eq.~(\ref{localham}) with photo-detection (d), a highly entangled state is stabilised for almost all parameter space.} 
\label{figureSE}
\end{figure}

Let us consider first the scheme with symmetric control where, in the absence of atomic decay, symmetric and anti-symmetric subspaces were decoupled and the system presented infinitely many stationary solutions. Spontaneous emission, however, introduces a coupling between the subspaces as it pushes the system towards its ground state. During this process, the anti-symmetric component $\ket{4}$ is produced, breaking the symmetry of the system. This has a huge impact on the new steady state of the system, which is now unique and, for most of the parameters, has no resemblance to the corresponding state for $\gamma=0$. The corresponding steady state entanglement, shown in Fig.~\ref{figureSE}c, is drastically reduced as compared to the situation of Fig.~\ref{fig3}b, despite the small value of the atomic decay rate. 
  
The scenario changes drastically when the local feedback Hamiltonian~(\ref{localham}) is considered. In this case, even without spontaneous decay, the system could explore the whole Hilbert space as the original symmetry was already broken by the control Hamiltonian. Moreover, the steady states, for both homodyne and jump-based feedback, were unique and highly entangled. The results for the steady state entanglement for  $\gamma=0.01 \, \Gamma$ are shown in Figs.~\ref{figureSE}b and \ref{figureSE}d for homodyne and jump feedback, respectively. Differently form the $J_x$ control case, spontaneous emission has indeed only a perturbative effect, reducing slightly the maximum value of steady state entanglement (compare with Fig.~\ref{figurelocal}). The figure also indicates that, for the local jump control,  the steady state remains close to the anti-symmetric Bell state for most values of $\lambda$ and $\Omega$, preserving the plateau structure shown in Fig.~\ref{figurelocal}b (see Fig.~\ref{figureSE}d). In fact, this can be shown directly  from the analytical solution including atomic decay.  The full solution is complicated but, expanding in powers of $\gamma/\Gamma$ and keeping only the lowest order term, one obtains 
\bea
\label{sss_jumplocal}
\rho_{44} &\approx& 1-O(\gamma/\Gamma), \nonumber \\
\rho_{ij} &\approx& O(\gamma/\Gamma), \: i,j\ne 4.
\eea
The final state is therefore close to the anti-symmetric Bell state. The first order terms depend on the parameters $\Omega$ and $\tilde \lambda$, and this dependence is responsible for the small deviations from the plateau structure of Fig.~\ref{figurelocal}c.

The major reason for this robustness of a local control strategy using both detection schemes lies in the uniqueness of the steady state without decay. For any initial condition, the system is driven to the target state and, therefore,  as soon as spontaneous decay forces the system away from this state, feedback dynamics counteract this tendency. This competition between feedback and decay dynamics determines how far the perturbed stationary state will be from the original one. Consequently, as soon as the ratio between atomic and collective decay rates remain small, the final state will remain highly entangled.

These results are not restricted to a feedback Hamiltonian of the form Eq.~(\ref{localham}). In fact, similar steady state entanglement are obtained other local control laws, though less entanglement can also be observed for particular regions of parameters. Therefore, there is still room for improvements in the results shown in Fig.~\ref{figureSE} as one can, in principle, optimise over the forms of $U$ to get even larger final entanglement.

\subsection{Detection inefficiencies effects}

Since feedback relies on the manipulation of the system based on information gained by a measurement, evaluation of the effects of inefficiencies in the detection process is important. 
Finite efficiency can be introduced both for photo-detection~\cite{wise_milb_jump93} and homodyne measurements~\cite{wise_milb_homo93}, and this has been explicitly considered for homodyne feedback schemes~\cite{wise_milb93,wiseman02,wang01}. From now on, we will consider only the jump-based feedback, since it is the scheme with best performance for entanglement generation. In this case, the extension of Eq.~(\ref{eq:pd}) to allow for a non-unit-efficiency detection can be done by identifying two distinct situations when a jump occurs: in the first the detector clicks and the feedback transformation $U$ acts in the system, in the second the detector fails to click and no control is applied. The corresponding equation reads 
\bea
\label{eq:pd_eta}
\dot \rho=-\frac{i}{\hbar}\Omega\left[(J_+ + J_-),\rho \right] + \Gamma \eta {\cal D}[U J_-]\rho \nonumber \\+ \Gamma \left(1-\eta\right) {\cal D}[J_-]\rho  + \sum_i \gamma_i{\cal D}[\sigma_i]\rho .
\eea
When the detector efficiency $\eta$ is zero, no information is extracted from the measurement and the equation reduces to the equation without feedback, Eq.~(\ref{eq:me_dicke}). Evidently, for an unity efficiency detector Eq.~(\ref{eq:pd}) is regained, and, for a local control, a maximally entangled steady state is reached. In the intermediate case where $0<\eta<1$, one would expect that imperfect knowledge gain should lead to inefficient control. Note however that, neglecting spontaneous emission, the anti-symmetric Bell state is a steady state of both Eqs.~(\ref{eq:me_dicke}) and~(\ref{eq:pd}), and one can show, proceeding exactly as in the unit-efficiency case, that this also holds true for Eq.~(\ref{eq:pd_eta}) for any $\eta >0$. This is a peculiar situation of this dynamics: while the feedback term  tries to move the system to the state $\ket{4}$, the one corresponding to a missed click does not affect this anti-symmetric component, impinging only a delay in the time taken to reach the steady state~\cite{jumpfeedback}.

\begin{figure}
\includegraphics[width=8.0cm]{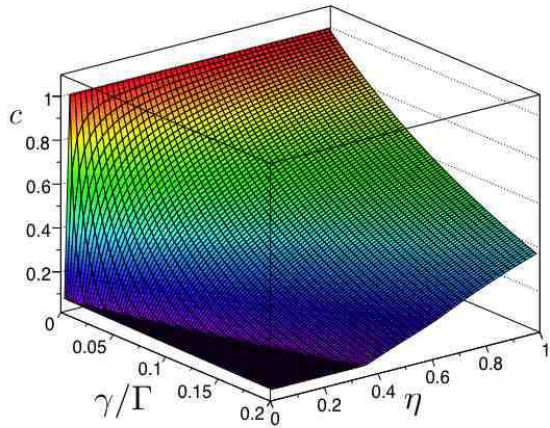}
\caption{(Color online) Maximum steady state concurrence as a function of the ratio $\gamma/\Gamma$ and detection efficiency $\eta$. For small $\gamma/\Gamma$ the system is almost insensitive to detection inefficiencies, decaying abruptly when $\eta$ approaches zero (for $\gamma/\Gamma=0$ entanglement is maximum unless $\eta=0$).}
\label{figetagamma}
\end{figure}

This is not true if atomic decay is taken into account as undetected events imply missed opportunities to apply the feedback, hence an effectively weaker feedback as compared to spontaneous emission. Mathematically this is clear from Eq.~(\ref{eq:pd_eta}) as the feedback term is scaled by $\eta$. The effect of detector inefficiencies and spontaneous emission together are summarised in Fig.~\ref{figetagamma} where the maximum steady state concurrence is plotted as a function of $\eta$ and $\gamma/\Gamma$. The most noticeable feature is the weak dependence of the entanglement on the detection efficiency in the limit $\gamma \ll \Gamma$ (unless $\eta$ is close to zero) that confirms the robustness of a local jump-based feedback in this regime~\cite{jumpfeedback}. For a ratio $\gamma/\Gamma=0.002$, for example, one still has concurrence above $0.9$ for detection efficiency as low as $\eta=0.1$

\section{Analysis of the adiabatic approximation}
\label{sec:exp}

The analysis from the previous sections shows that entanglement production can be vastly improved by the use of feedback strategies and that, in the best case scenario, highly entangled states can be protected against decoherence using a quantum-jump-based feedback with a local control Hamiltonian. The robustness of the feedback scheme depends on a single quantity, the ratio $C=\Gamma/\gamma=g^2/\kappa \gamma$, also known as the cooperativity parameter~\cite{bonifacio78}.
The higher the cooperativity, the better the performance of the feedback procedure: not only does the steady state entanglement obtained increase, but also the sensitivity of the scheme to detection inefficiencies is reduced. 

Interestingly, large cooperativity, which corresponds to the strong coupling limit, is the regime used by many of the current cavity QED experiments in the optical domain~\cite{hood00,puppe07,sauer04}. This is achieved either by increasing the atom-cavity coupling or decreasing the cavity decay rate, since the spontaneous emission rate is fixed by the choice of atomic species. The use of high quality cavities, however, affects the feedback model based on Eq.~(\ref{eq:me_total}), which is obtained in the bad cavity limit after adiabatic eliminating the cavity mode. To have a better understanding of how our model would work within current experimental conditions and also to determine the most favourable region of parameters for future experiments, in this section we will relax the adiabatic condition and analyse the performance of local jump feedback strategy in this situation.

\subsection{The non-adiabatic model}\label{sec:nonadiabaticmodel}
  Equation~(\ref{eq:nofb_full}) describes the whole system, including cavity mode dynamics, without control. Introduction of photo-detection feedback follows exactly the same reasoning used to obtain Eq.~(\ref{eq:pd}). Whenever a photon is detected, the control Hamiltonian is applied and the system undergoes an evolution given by the operator $U$. This operator now enters in the term that describes the monitoring of the cavity output, {\it i.e.} ${\cal D}[a] \rho$. Addition of detection inefficiency also follows directly from our previous discussion and the full master equation for the system is 
\begin{eqnarray}
\label{eq:fullfeedback}
\dot \rho=-i \Omega \left[(J_+ + J_-),\rho \right] -i g \left[(J_+ a + J_- a^\dagger),\rho \right] \nonumber \\  + \eta \kappa {\cal D}[Ua]\rho + (1 - \eta ) \kappa {\cal D}[a]\rho + \sum_i\gamma_i{\cal D}[\sigma_i]\rho,
\end{eqnarray}
which transforms back to Eq.~(\ref{eq:pd_eta}) if an adiabatic elimination is performed.

To start with, instead of solving the full master equation, Eq.~(\ref{eq:fullfeedback}), we will use a quantum trajectory method~\cite{carmichael,molmer96} to follow the dynamics conditioned on the measurement results. This approach, which provides a useful picture of single runs of an actual experiment, combines continuous evolution under an effective Hamiltonian and the random occurrence of quantum jumps. In our problem there are four possible jumps, corresponding to the four decoherence terms in Eq.~(\ref{eq:fullfeedback}): a detected photon leaving the cavity that triggers the feedback (third term), an undetected photon (fourth term), and spontaneous emission from the atoms (two jumps represented by the sum in the fifth term). For simulation purposes, we assume that the environment is perfectly monitored so that the state remains pure throughout the evolution. Obviously this is not true for undetected photons, but we can take a different perspective and consider feedback inefficiencies as detected photons where the feedback Hamiltonian fails to apply. Note that for the average behavior given by Eq.~(\ref{eq:fullfeedback}) those points of view are equivalent.

\begin{figure}
\includegraphics[width=8.5cm]{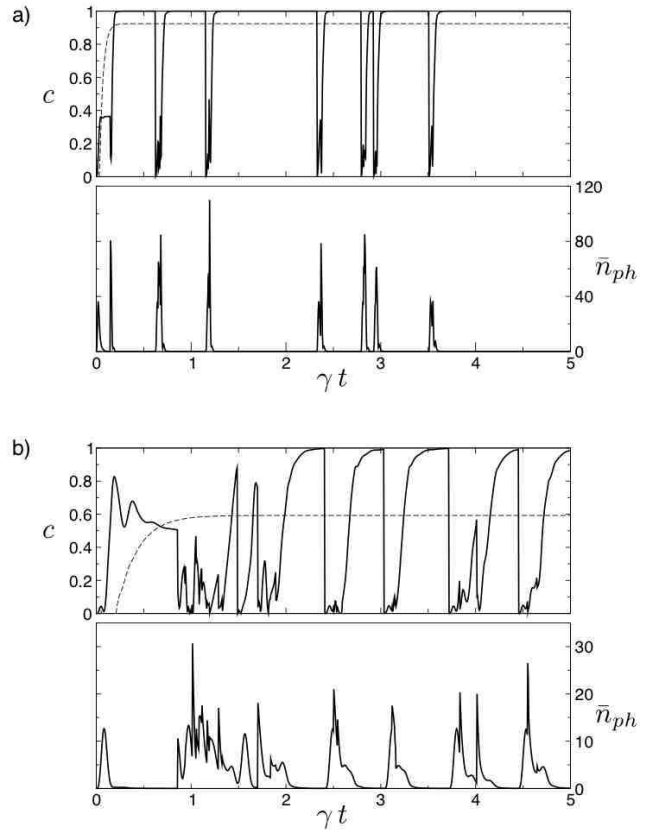}
\caption{Concurrence (top panels) and mean cavity photon number (bottom panels) as a function of time for single quantum trajectories. In (a) the system is in the adiabatic regime with $\kappa=400 \gamma$,  $g=200 \gamma$, $\Omega=40 \gamma$ and $\tilde \lambda=-18 \gamma$. In (b), effects of non-adiabatic dynamics get more evident ($\kappa= 16\gamma$,  $g=40 \gamma$, $\Omega=20.8 \gamma$ and $\tilde \lambda=-10.0528 \gamma$). In both plots, the cooperativity is $C=100$, and the concurrence for the time-evolved density matrix is shown for comparison (dashed lines)}
\label{singleshots}
\end{figure}

Figure~\ref{singleshots} shows the concurrence (top panels) and the average number of photons in the cavity (bottom panels) for single stochastic trajectories as a function of time. The cooperativity parameter is set to $C=100$ (as in Fig.~\ref{figureSE}), the detection efficiency is $\eta=1$, and we assume that the atoms are initially in the ground state and the cavity in the vacuum. Figure~\ref{singleshots}a corresponds to a situation close to the adiabatic regime with $\kappa=400 \gamma$ and $g=200 \gamma$,  while $g=40 \gamma$ and $\kappa=16 \gamma$ in Fig.~\ref{singleshots}b. All other parameters are selected to give the maximum concurrence for the chosen $g$ and $\kappa$. The average behavior, given by the solution of Eq.~\ref{eq:fullfeedback}, is shown by the dashed lines for comparison.

For the case of Fig.~\ref{singleshots}a, dynamics with feedback drives the system to the target state in a short time. In this particular realization, two detected photons, and consequently two control pulses, at $t\approx 0.15$ are enough to correct the trajectory to state $\ket{4}$ with no photons in the cavity. The absence of photons from the cavity output for certain periods of time is therefore an indication that the atoms are in the anti-symmetric Bell state. This state can only be perturbed by the occurrence of spontaneous emission jump (in this case at $\gamma\,t\approx 0.62, 1.15, 2.3, 2.8, 2.9, \,{\rm and} \, 3.5$), which destroys entanglement. As soon as the system gets away from the dark state, cavity is again populated and control pulses applied when photons are detected, reestablishing the maximally entangled state. In the adiabatic regime, since $\kappa \gg \gamma$, this happens quickly during short windows of control jumps (we have 7 of them in Fig.~\ref{singleshots}a). As a result, the system spends most of the time in the entangled state and, on average, entanglement will be high. This is another way of understanding the perturbative nature of the spontaneous emission in the adiabatic limit: the larger the cooperativity, the larger the amount of time spent in the target state, hence higher entanglement. 

However, this picture is only valid in the adiabatic limit where cooperativity is the only important parameter. Fig.~\ref{singleshots}b shows the dynamics outside the adiabatic regime but for the {\it same} cooperativity. Now, with a smaller $\kappa$, the windows of control are larger and the system takes longer to reach the target state. Moreover, the ratio between the probabilities of having spontaneous emission and control events increases. Consequently, the system spends less time in the state $\ket{4}$, resulting in a less entangled steady state ($c \approx 0.6$). This may be explained by the reliance of the control scheme on information transmitted by the photons that leak from the cavity.  In the same way that low detection efficiency limits the eventual peak concurrence because an opportunity for control is lost, if the cavity decay rate is very low, the photons that are emitted into the cavity have a higher probability of being reabsorbed by the atoms rather than escaping and triggering a control pulse.

One can therefore conclude that the major aspect responsible for the robustness of the local jump-based feedback scheme is the balance between control and decay dynamics. In the adiabatic case, the feedback time scale is dictated by the effective rate $\Gamma$ and the robustness depends only on the cooperativity $\Gamma/\gamma$. In the non-adiabatic case, feedback rate is given by the cavity decay $\kappa$ and the feedback performance will crucially depend on the ratio $\kappa/\gamma$. This indicates that the adiabatic regime is the best one to achieve a robust generation of highly entangled states. 

\begin{figure}
\includegraphics[width=8.0cm]{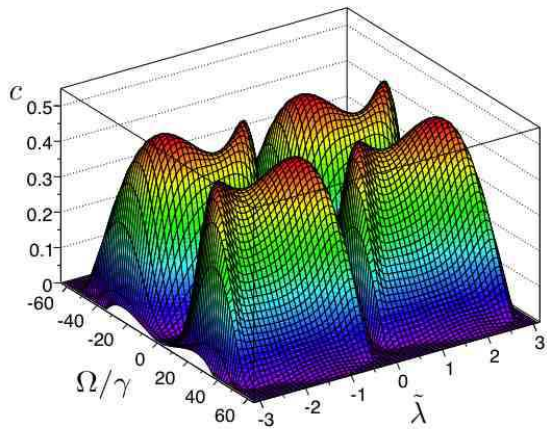}
\caption{(Color online) Steady state concurrence for jump-based feedback as a function of driving and feedback strengths for $g=40 \gamma$,$\kappa = 16 \gamma$, and $\eta = 0.5$. The maximum concurrence ($c\approx 0.5$) occurs for specific values of the parameters in contrast with the flat structure observed in the adiabatic regime (see Fig.~\ref{figureSE}d). }
\label{fig_structures}
\end{figure}

Note, however, that the ratio $\kappa/\gamma$ is not the only important parameter in the non-adiabatic case. The laser can excite the atoms, which can then emit photons in the cavity mode. These photons can finally escape the cavity, triggering the feedback. The dynamics depends, in an intricate way, on how the rates at which those process occur are related, contributing significantly for the behavior of the steady state entanglement. Figure~\ref{fig_structures} shows the stationary concurrence as a function of $\Omega$ and $\tilde \lambda$ for the same values of $g$ and $\kappa$ as in Fig.~\ref{singleshots}b, but for a smaller detection efficiency $\eta=0.5$. The simple plateau feature from the adiabatic case (Fig.~\ref{figureSE}d) disappears, giving place to a structure that highlights the strong dependence of the final entanglement on the different parameters of the problem.   

\section{Conclusions}\label{sec:conc}
In this paper we have investigated the improvement on the generation of steady state entanglement via quantum Markovian feedback. Using the Dicke model, we have explored the influences of the choice of detection schemes to the success of the control, showing that a photo-detection feedback strategy outperforms the one based on homodyne measurements. The effects of different feedback Hamiltonians were also analysed: local control schemes, {\it i.e} feedback applied to only one of the atoms, produced larger amounts of steady state entanglement than collective interaction. We identified the strategy of jump-based feedback with local control~\cite{jumpfeedback} to be the best for robust preparation of highly entangled states. For this scheme, we extended the numerical analysis given in~\cite{jumpfeedback}, and presented a justification for the robustness against spontaneous emission and detection inefficiencies based on analytical considerations. Motivated by the parameters used in most of the recent optical cavity QED experiments, we also investigated the role of the adiabatic elimination on our feedback scheme and showed that the adiabatic regime gives the best performance in terms of entanglement generation and robustness. 

An important issue is whether our feedback could be realized experimentally. The setup of two atoms equally coupled to a cavity mode with possibility of individual addressing has already been demonstrated in~\cite{nussmann_05}. The remaining constraints for an efficient feedback procedure are: i) large cooperativity parameter  ($\Gamma \gg \gamma$) and ii) high detection rate as compared to the decay rate ($\kappa \gg \gamma$). Although the later condition can be certainly achieved in experiments with low quality cavities, and the former has been obtained in a variety of recent experiments~\cite{hood01,khudaverdyan08,boozer06,maunz05}, the difficulty relies on combining all those ingredients in a single experimental setup. The usual strategy to fulfill (i) is to improve the quality of the cavity, which would be detrimental to condition (ii). The solution would be to design a cavity with high transmissivity (ii), yet with a large coupling strength $g$, which would involve a compromise between small mode volume and the requirement of individual addressing of atoms.

\begin{acknowledgments}
The authors thank M. Hush for helpful discussions and comments on the manuscript. A. R. R. C. also thanks D. Meschede for enlightening discussions about experimental issues.  
\end{acknowledgments}

\label{sec_concl}

\bibliography{qcontrol}

\end{document}